# Spherical Geometry Algorithm for Space-borne Synthetic Aperture Radar Imaging

Xinhua Mao

*Abstract*—Higher spatial resolution and larger imaging scene are always the goals pursued by advanced spaceborne SAR system. High resolution and wide swath SAR imaging can provide more information about the illuminated scene of interest on one hand, but also come with some new challenges on the other hand. The induced new challenging problems include curved orbit, Earth rotation, and spherical ground surface, etc. Most existing image formation algorithms suffer from performance deficiency in these challenging cases, either in focus accuracy or computational efficiency. In this paper, an accurate Fourier transform relationship between the phase history domain data and the scene reflectivity function is derived under arbitrary radar trajectory by exploiting the spherical geometry property of the spaceborne SAR data collection. Using the derived new data model, an image reconstruction algorithm based on Fourier inversion is proposed. The new algorithm has the inherent capability of correcting for the curved orbit and spherical ground surface effect. Meanwhile, the out-of-plane motion effect induced by the Earth's rotation can also be compensated by a two-step phase correction and data projection procedure embedded in the Fourier inversion reconstruction. The new algorithm inherits the merit of both time domain and frequency domain algorithms, has excellent performance in both focus accuracy and computational efficiency. Both simulation and real data processing results validate the effectiveness of the proposed imaging algorithm.

*Index Terms*—synthetic aperture radar, high resolution and wide swath, spherical ground surface, nonlinear trajectory, spherical geometry algorithm.

## I. INTRODUCTION

SPACEBORNE synthetic aperture radar (SAR) is an important remote sensing tool due to its all day and all weather imaging capability[1]. As an imaging radar, spatial resolution and imaging scene size are two important indexes that determine the ability of radar to obtain information from the illuminated scene. Therefore, higher spatial resolution and larger imaging scene are always the goals pursued by advanced spaceborne SAR system. At present, the highest resolution of spaceborne SAR in orbit has reached the submeter level, e.g., TerraSAR has 0.16m resolution in azimuth and 0.5m in range direction [2-3]. In the near future, the resolution will get close to decimeter or even centimeter level. As for the scene width, typical spaceborne SAR imaging scene width is in the order of tens of kilometers, but now there are also some advanced SAR system whose imaging swath width have reached hundreds of kilometers or even larger, especially for future MEO or GEO SAR [4-5]. The continuous improvement of resolution and scene width enriches the information obtained by radar, but on the other hand, it also brings great challenges to the SAR system implementation and image formation processing. As far as the image formation processing is concerned, the induced new challenging problems include curved orbit, Earth rotation, and spherical Earth surface. On the one hand, higher resolution means longer synthetic aperture time [6-9]. During a considerable synthetic aperture time, the radar flight path is inevitably curved due to the inherent circular or elliptical satellite orbit, as shown in Fig.1(a). To make matters worse, when considering the Earth rotation, radar trajectory even become a much more complex non-coplanar curve in the Earth Center Earth Fixed (ECEF) coordinate frame, as shown in Fig.1(b). On the other hand, when the illuminated scene become very large, as shown in Fig.1(c), the flat ground assumption is also no longer reasonable. An accurate image formation algorithm must take into account the spherical ground surface effect.

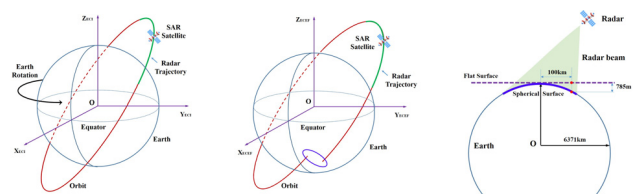

(a) curved flight path  (b) non-coplane flight path  (c) spherical ground surface

**Fig. 1.** Three challenging problems for high resolution and wide swath spaceborne SAR imaging.

In the spaceborne SAR literature, many efforts have been made to find accurate and efficient image formation algorithms to overcome these problems. The proposed algorithms can be divided into two categories, i.e., time-domain algorithms and frequency-domain algorithms. Time domain algorithms are very accurate even under very complex imaging scenarios, because there is no any approximations during its formulation [10-11]. In addition, because they perform phase adjustment and correction on the echo signal in azimuth time domain for each pixel independently, they have an inherent capability of correcting for any nonlinear radar platform motion and any scene topography. Therefore, if only considering the focus accuracy, time domain algorithms are the best choice for high resolution wide scene spaceborne SAR image formation processing. Nevertheless, time domain algorithms also have a fatal weakness, i.e., very high computational complexity, which limits their applications in practical systems. To improve computational efficiency, some new modified algorithms are proposed, but all at the expense of focus accuracy and flexibility [12-13]. Furthermore, for time domain algorithm, when accurate flight trajectory information cannot be available, it is often very difficult to be compatible with some efficient autofocus approach, such as PGA, MD [14-15]. On the contrary, frequency domain algorithms possess high computational efficiency because they use fast Fourier transform (FFT) to realize batch processing. They are almost always the preferred option for practical imaging processor. However, conventional frequency domain algorithms used in spaceborne SAR, such as RD, CS, RMA, originally all come from airborne SAR applications. These algorithms have two common assumptions that the radar flight trajectory during the synthetic aperture time is linear and the illuminated scene is planar. Under these assumptions, the range history between radar and target can be modeled as a hyperbolic curve which can greatly facilitate the high efficient image formation processing. However, for very high resolution spaceborne SAR, the actual range history often deviates far from the hyperbolic model due to the nonlinear radar trajectory resulted from the curved orbit and Earth rotation [16-17]. In order to model the range history more accurately in the high-resolution case, many improved range models are proposed. The basic idea of these new models is to increase the order of the Taylor expansion approximation of the range history [18-23]. Although these new models can improve the algorithm accuracy greatly, frequency domain algorithm based on these models still suffer from at least two shortcomings which limit their application in ultra-high resolution SAR. First, limited low-order Taylor expansion approximation may still cannot satisfy the accuracy requirement of very-high-resolution SAR imaging. Secondly, space-variant phase history is very sensitive to ground topography when the synthetic aperture is very long [24-25]. Therefore, it is still a very difficult task to eliminate the space-variant effect through batch processing for frequency domain algorithm, especially when the spherical ground surface effect is taken into account. In a nutshell, there are almost no algorithms that can absolutely meet the requirements of the focus accuracy and computational efficiency at the same time for very high resolution spaceborne SAR imaging of very large scene.

In this paper, different from some conventional approaches that regard the spherical ground surface as an adverse factor in the image formation processing, we take it as a favorable prerequisite. By exploiting the spherical ground surface geometry of the spaceborne SAR data collection, an accurate Fourier transform relationship between the phase history domain data and the scene reflectivity function is derived. Using the derived new data model, an image reconstruction algorithm based on Fourier inversion is proposed. The new algorithm has the inherent capability of correcting for the curved orbit and spherical ground surface effect. Meanwhile, the out-of-plane motion effect induced by the Earth's rotation can also be compensated by a two-step phase correction and data projection operation embedded in the Fourier reconstruction procedure. Therefore, the new algorithm inherits the merit of both time domain and frequency domain algorithms, has excellent performance in both focus accuracy and computational efficiency.

The rest of the paper is organized as follows. In section II, the Fourier relationship between phase history data and illuminated scene function is derived by exploiting the spherical geometry property of the spaceborne SAR imaging. Using the new model, a Fourier reconstruction approach is proposed for low resolution spaceborne SAR in section III. In section IV, to eliminate the Earth rotation effect in high resolution spaceborne SAR, a modified Fourier reconstruction approach is proposed. Finally, in section V, experimental results are presented to demonstrate the effectiveness of the proposed approach.

## II. FOURIER RELATIONSHIP BETWEEN DATA AND TARGET FUNCTION

In this section, we will formulate the signal model for spaceborne SAR and reveal the Fourier transform relationship between the collected data (after some preprocessing) and the illuminated scene function. This analytical relationship will provide a prerequisite for the proposed image formation algorithm in the following sections.

### A. Imaging Geometry

To describe the satellite orbit, an Earth Centered Inertial (ECI) coordinate system is often preferred because in this frame the trajectory of the satellite is a standard circle or ellipse in a plane. However, in SAR image formation processing, what we are concerned about is the relative geometric relationship between the satellite and the Earth. Therefore, it will be more convenient to use Earth Centered Earth Fixed (ECEF) coordinate system. However, due to Earth's rotation, the trajectory of the satellite in the ECEF frame will become a non-coplanar curve, as shown in Fig.1(b).

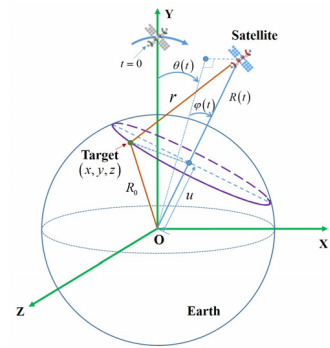

**Fig. 2.** Data collection geometry in rotated ECEF coordinate system.

Without loss of generality, the imaging geometry of a spaceborne SAR system is shown in Fig.2. To further facilitate the SAR signal analysis and image formation processing, a rotated ECEF coordinate system is adopted. In this coordinate system, the center of the Earth is defined as the origin, the Y axis is defined to point towards the SAR location at the aperture center, the X axis is defined in the plane determined by the Y axis and the radar velocity vector at the aperture center, and the Z axis is orthogonal to the X and Y axes with the right-hand rule. Let $t$ represent the slow time and define $t = 0$ at the aperture center. The instantaneous distance from the Earth center to the radar is denoted as $R(t)$. The instantaneous azimuthal angle and grazing angle of the radar with respect to the Earth center are denoted as $\theta(t)$ and $\varphi(t)$, respectively. Therefore, the instantaneous position of the radar platform can be expressed as $(R,\theta,\varphi)$ in spherical coordinate or $(R\cos\varphi\cos\theta, R\cos\varphi\sin\theta, R\sin\varphi)$ in Cartesian coordinate. To simplify notations, we omit the azimuth time dependent effect when there is no risk of confusion. For a general point located at $(x,y,z)$ in the 3-D space, the instantaneous range from this point to the radar can be expressed as

$$r=\sqrt{(x-R\cos\varphi\sin\theta)^2+(y-R\cos\varphi\cos\theta)^2+(z-R\sin\varphi)^2} . \quad (1)$$

For a fixed range $r$ and radar position $(R,\theta,\varphi)$, this equation defines a sphere centered at the SAR location with a radius of $r$, and also describes accurately the spherical wavefront of the radiated electromagnetic waves. Radar echo signals from all the scatterers in this sphere will arrive at the radar receiver simultaneously, therefore, the received signal is a superposition of echo signal scattered from all the targets on the sphere. In actual, the scatterers of interest only distribute on the surface of the Earth. The intersection of the Earth surface with the constant-range sphere centered at the radar position is a circle which lies in a plane orthogonal to the SAR position vector. This plane can be described by

$$x\cos\varphi\sin\theta + y\cos\varphi\cos\theta + z\sin\varphi = u , \quad (2)$$

where $u$ is the perpendicular distance from the Earth center to the plane. By inserting the Earth surface equation ($x^2+y^2+z^2=R_0^2$, $R_0$ is the radius of the Earth) and (2) into (1), it is easily to obtain the relationship between $u$ and $r$

$$r=\sqrt{R^2+R_0^2-2Ru} , \quad \text{or} \quad u = \frac{R^2+R_0^2-r^2}{2R} . \quad (3)$$

These equations can also be obtained from the geometry in Fig.2 by using the law of cosines.

*B. Echo Signal Model*

Assuming that the radar mounted on a satellite transmits a wideband signal modulated at a carrier frequency $f_c$ given by

$$s_c(\tau) = s_b(\tau) \cdot \exp(j2\pi f_c \tau), \quad (4)$$

where $s_b(\tau)$ is the baseband signal, and $f_c$ is the carrier frequency.

Because all the scatterers lying along the same constant-range contour will be received by the radar at precisely the same time, the received signal at a particular time is the returns from the integration of the reflectivity values from all targets that lies along the corresponding constant range sphere determined by (1). If we assume that the function $g(x,y,z)$ represents the three-dimensional radar reflectivity density function, then the integration function can be expressed as

$$p_{\theta,\varphi}(r) = \iiint g(x,y,z) \cdot \delta(r-r_{x,y,z}) dxdydz , \quad (5)$$

where $r_{x,y,z} = \sqrt{(x-R\cos\varphi\sin\theta)^2+(y-R\cos\varphi\cos\theta)^2+(z-R\sin\varphi)^2}$,

and $\delta(\cdot)$ is the Dirac delta function.

We neglect the atmospheric effect and any motion during the transmission and reception, then the expression for the returned radar signal at the receiver is given by

$$\begin{aligned} S(t,\tau) &= \int_r p_{\theta,\varphi}(r) \cdot s_c\left(\tau - \frac{2r}{c}\right) dr \\ &= \int_r p_{\theta,\varphi}(r) \cdot s_b\left(\tau - \frac{2r}{c}\right) \cdot \exp\left\{j2\pi f_c\left(\tau - \frac{2r}{c}\right)\right\} dr \end{aligned} . \quad (6)$$

*C. Fourier Transform Relationship*

To show the Fourier transform relationship between the collected data and the target function, we need to perform some preprocessing on the echo signal.

First, a pulse compression processing is performed in the range direction. After pulse compression, the point spread function can often be approximated as a sinc function. Therefore, the range compressed echo signal can be expressed as

$$S(t,\tau) = \int_r p_{\theta,\varphi}(r) \cdot \text{sinc}\left(B_r\left(\tau - \frac{2r}{c}\right)\right) \cdot \exp\left\{j2\pi f_c\left(\tau - \frac{2r}{c}\right)\right\} dr . \quad (7)$$

where $B_r$ is the bandwidth of the transmitted signal.

Using the property of Dirac delta function [26], the delta function in (5) can also be expressed as

$$\begin{aligned} &\delta\left(r - \sqrt{(x-R\cos\varphi\sin\theta)^2+(y-R\cos\varphi\cos\theta)^2+(z-R\sin\varphi)^2}\right) \\ &= \alpha \cdot \delta\left(\frac{R^2+x^2+y^2+z^2-r^2}{2R} - (x\cos\varphi\sin\theta + y\cos\varphi\cos\theta + z\sin\varphi)\right) \end{aligned} , \quad (8)$$

where $\alpha$ is a constant.

Because we only need to take into account the scatterers distributed on the Earth surface, the coordinate of the scatterers satisfies $x^2+y^2+z^2 = R_0^2$. Then, equation (8) can be rewritten as

$$\begin{aligned} &\delta\left(r - \sqrt{(x-R\cos\varphi\sin\theta)^2+(y-R\cos\varphi\cos\theta)^2+(z-R\sin\varphi)^2}\right) \\ &= \alpha \cdot \delta\left(\frac{R^2+R_0^2-r^2}{2R} - (x\cos\varphi\sin\theta + y\cos\varphi\cos\theta + z\sin\varphi)\right) \end{aligned} . \quad (9)$$

Using (3), (9) can also be rewritten as

$$\begin{aligned} &\delta\left(r - \sqrt{(x-R\cos\varphi\sin\theta)^2+(y-R\cos\varphi\cos\theta)^2+(z-R\sin\varphi)^2}\right) \\ &= \alpha \cdot \delta(u - x\cos\varphi\sin\theta - y\cos\varphi\cos\theta - z\sin\varphi) \end{aligned} . \quad (10)$$

Then the integration function in (5) can be simplified as

$$\begin{aligned} p_{\theta,\varphi}(r) = \alpha \cdot \iiint g(x,y,z) \\ \cdot \delta(u - x\cos\varphi\sin\theta - y\cos\varphi\cos\theta - z\sin\varphi) dxdydz \end{aligned} . \quad (11)$$

Using (11), the range compressed signal in (7) can be rewritten as

$$\begin{aligned} S(t,\tau) = \int_u \iiint_{x,y,z} g(x,y,z) \cdot \delta(u - x\cos\varphi\sin\theta - y\cos\varphi\cos\theta - z\sin\varphi) \\ \cdot \text{sinc}\left(B_r\left(\tau - \frac{2r}{c}\right)\right) \cdot \exp\left\{j2\pi f_c\left(\tau - \frac{2r}{c}\right)\right\} dxdydzdu \end{aligned} . \quad (12)$$

where the nonessential amplitude factors are ignored for clarity.

Then, we define a change of variable which maps a new range time variable $\tau'$ into the range time $\tau$ by

$$\tau = \zeta(\tau') = \frac{2}{c}\sqrt{R^2 + R_0^2 - Rc\tau'} \quad . \tag{13}$$

By exploiting (3), this map has the property: $\zeta\left(\frac{2u}{c}\right) = \frac{2r}{c}$.

After this change of variable, the echo signal becomes

$$S(t,\tau') = \int\limits_u \iiint\limits_{x,y,z} g(x,y,z) \cdot \delta(u - x\cos\varphi\sin\theta - y\cos\varphi\cos\theta - z\sin\varphi)$$
$$\cdot \mathrm{sinc}\left(B_r\left(\zeta(\tau') - \frac{2r}{c}\right)\right) \cdot \exp\left\{j2\pi f_c\left(\zeta(\tau') - \frac{2r}{c}\right)\right\} dxdydzdu \quad . \tag{14}$$

Because the sinc function can be approximated as a local function, whose energy is distributed mainly in the vicinity of its maximum value, function $\zeta(\tau')$ can then be approximated as a linear function in the vicinity of $\frac{2u}{c}$

$$\zeta(\tau') \approx \zeta\left(\frac{2u}{c}\right) + a\left(\tau' - \frac{2u}{c}\right) = \frac{2r}{c} + a\left(\tau' - \frac{2u}{c}\right), \tag{15}$$

where $a = \left.\frac{d\zeta(\tau')}{d\tau'}\right|_{\tau' = \frac{2u}{c}} = -\frac{R}{r}$.

Inserting (15) into (14), we can obtain

$$S(t,\tau') = \int\limits_u \iiint\limits_{x,y,z} g(x,y,z) \cdot \delta(u - x\cos\varphi\sin\theta - y\cos\varphi\cos\theta - z\sin\varphi)$$
$$\cdot \mathrm{sinc}\left(\bar{B}_r\left(\tau' - \frac{2u}{c}\right)\right) \cdot \exp\left\{j2\pi \bar{f}_c\left(\tau' - \frac{2u}{c}\right)\right\} dxdydzdu \quad , \tag{16}$$

where $\bar{B}_r = aB_r$ and $\bar{f}_c = af_c$.

After removing the carrier frequency, the demodulated signal becomes

$$S(t,\tau') = \int\limits_u \iiint\limits_{x,y,z} g(x,y,z) \cdot \delta(u - x\cos\varphi\sin\theta - y\cos\varphi\cos\theta - z\sin\varphi)$$
$$\cdot \mathrm{sinc}\left(\bar{B}_r\left(\tau' - \frac{2u}{c}\right)\right) \cdot \exp\left\{-j\frac{4\pi \bar{f}_c}{c} u\right\} dxdydzdu \quad . \tag{17}$$

Using the integration property of Dirac delta function, (17) can also be expressed as

$$S(t,\tau') = \iiint\limits_{x,y,z} g(x,y,z) \cdot \mathrm{sinc}\left(\bar{B}_r\left(\tau' - \frac{2}{c}(x\cos\varphi\sin\theta + y\cos\varphi\cos\theta + z\sin\varphi)\right)\right)$$
$$\cdot \exp\left\{-j\frac{4\pi \bar{f}_c}{c}(x\cos\varphi\sin\theta + y\cos\varphi\cos\theta + z\sin\varphi)\right\} dxdydz \quad . \tag{18}$$

Finally, performing a range Fourier transform on (18) yields

$$S(t,f_\tau) = \iiint\limits_{x,y,z} g(x,y,z) \cdot \exp\left\{-j\frac{4\pi}{c}(\bar{f}_c + f_\tau)\right.$$
$$\left. \cdot (x\cos\varphi\sin\theta + y\cos\varphi\cos\theta + z\sin\varphi)\right\} dxdydz \quad . \tag{19}$$
$$f_\tau \in \left[-\bar{B}_r/2, \bar{B}_r/2\right]$$

Equation (19) can also be rewritten as

$$S(t,f_\tau) = \iiint\limits_{x,y,z} g(x,y,z) \cdot \exp\left\{-jk_r(x\cos\varphi\sin\theta + y\cos\varphi\cos\theta + z\sin\varphi)\right\} dxdydz$$
$$k_r \in \left[\frac{4\pi}{c}(\bar{f}_c - \bar{B}_r/2), \frac{4\pi}{c}(\bar{f}_c + \bar{B}_r/2)\right] \quad , \tag{20}$$

where $k_r = \frac{4\pi}{c}(\bar{f}_c + f_\tau)$.

Define

$$k_x = k_r \cos\varphi\sin\theta$$
$$k_y = k_r \cos\varphi\cos\theta \quad , \tag{21}$$
$$k_z = k_r \sin\varphi$$

then, equation (20) can be rewritten as

$$S(t,f_\tau) = G(k_x, k_y, k_z), \tag{22}$$

where

$$G(k_x, k_y, k_z) = \iiint g(x,y,z) \cdot \exp\left\{-j(k_x x + k_y y + k_z z)\right\} dxdydz \quad , \tag{23}$$

is the 3-D Fourier transform of the target function $g(x,y,z)$.

Equation (22) shows clearly that, after some preprocessing, there is a Fourier transform relationship between the 2-D echo data and the illuminated target function, i.e., the echo data can be interpreted as discrete samples of the 3-D Fourier transform of the reflectivity of the illuminated scene. Thus, the imaging procedure is to generate an estimation of the target function from the collected Fourier data via an inverse discrete Fourier transform.

### D. Data Collection Manifold in Wavenumber Domain

From (20)-(23), it is clear that, after some preprocessing, the returned signal for a single pulse with the radar position of $(R,\theta,\varphi)$ is a slice at angle $(\theta,\varphi)$ of the 3-D Fourier transform of the unknown reflectivity function $g(x,y,z)$. The direction of the slice is determined by the two angles $\theta$ and $\varphi$ where the radar transmits and receives pulse signal, whereas the radial position and length of the slice are determined by the radar carrier frequency and bandwidth, respectively. As the radar platform moves through the synthetic aperture, a polar raster surface of the 3-D Fourier space will be swept out due to the variation of the angles $\theta$ and $\varphi$. Due to limited bandwidth of transmitted signal and limited observation angles, the collected echo data covers only a limited support area in the 3-D wavenumber domain. The precise shape of the swept collection surface is determined by the platform flight trajectory with respect to the Earth center. If the radar flight trajectory is a straight or curved line in a plane in the physical space, the data collection surface in the 3-D wavenumber space will also lies in a plane, as shown in Fig.3(a). However, if the radar flight trajectory is a non-coplanar curve, then the data collection surface will also become a curved surface, as shown in Fig.3(b). For spaceborne radar, the satellite orbit during the synthetic aperture time is a curved line in a plane in the inertial coordinate system. However, in the ECEF coordinate system, the satellite orbit will become a non-coplanar curve due to the rotation of the Earth. Nevertheless, for low resolution SAR, the non-coplane effect caused by the Earth's rotation can often be neglected because of the very small synthetic aperture time. In this case, the swept surface in the Fourier space can be approximated as a plane. However, for high resolution SAR, the out-of-plane motion due to the Earth's rotation will become commonplace. The resulted curved data collection surface in the wavenumber domain will greatly complicate the image formation processing implementation. In the following sections, we will discuss the image formation problems in these two cases respectively.

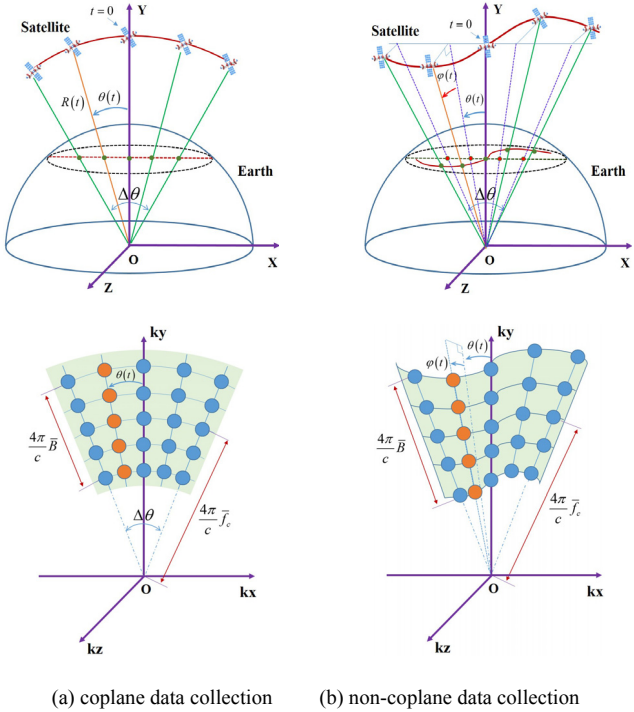

(a) coplane data collection  (b) non-coplane data collection

**Fig. 3.** Radar data collection geometry and its corresponding data manifold in wavenumber domain.

## III. SPHERICAL GEOMETRY ALGORITHM FOR LOW-RESOLUTION SAR

The goal of the SAR image formation processing is to estimate the target function from the gathered radar return data. In section II, we show the analytical relationship between the returned signal and the target function, which shows that, after some preprocessing, the processed echo data represents a slice in the 3-D Fourier domain of the target function. For a low-resolution SAR, the flight path of the radar platform can be approximated as a plannar curve because of the negligible effect of the Earth's rotation. In this case, the sampled Fourier data set lies in a plane. If we choose this plane as the focus plane, to reconstruct the target, all that is required is to inverse transform the sampled Fourier data into an image of the target by a 2-D inverse Fourier transform. From the discussion of the preceding section, we also know that the set of samples of Fourier domain lies on a polar raster imposed on an annulus. Therefore, in practical implementation of the image formation processing, to exploit the high efficiency of the 2-D IFFT to implement the 2-D inverse Fourier transform, a popular choice is to interpolate the 2-D polar-gridded samples into a rectangular grid for ultimate use in a 2-D IFFT image formation process.

Based on the above ideas, the proposed image formation algorithm for low-resolution spaceborne SAR includes three main processes. First, a preprocessing in the range dimension is performed on the demodulated raw data to produce the polar-gridded samples in the Fourier space. Then, a polar-to-rectangular resampling implemented by two separable 1-D resampling is exploited to interpolate the 2-D polar-gridded data into a rectangular grid data. Finally, a 2-D FFT or IFFT is applied to produce a well-focused imagery of the target.

### A. Range Preprocessing

From the previous section, we have already shown that the echo data after some preprocessing, instead of the raw data itself, are essentially a slice of the Fourier spectrum of the terrain reflectivity function. Therefore, to reconstruct the target function from the raw data by Fourier inversion, a preprocessing on the raw data is required firstly. The objective of the preprocessing process is to transform the 2-D raw signal shown in (6) into the phase history data shown in (19). To this end, four preprocessing steps are applied on the echo signal in the preceding derivation in section II. They are pulse compression, change-of-variable, demodulation, and Fourier transformation. However, the previous derivation assumes that the input signal for the image formation processor is an analog radio signal. In actual, the return signal is firstly demodulated into a baseband signal, and then the demodulated analog signal is transformed into digital signal by AD converter. We assume that the demodulated digital signal, which can be expressed as in (24), is the input data for the image formation processor.

$$S(t,\tau) = \iiiint_{r\,x,y,z} s_b\left(\tau - \frac{2r}{c}\right) \cdot \exp\left\{-j\frac{4\pi}{c}f_c r\right\} \cdot g(x,y,z)$$
$$\cdot \delta\left(r - \sqrt{(x-R\cos\varphi\sin\theta)^2 + (y-R\cos\varphi\cos\theta)^2 + (z-R\sin\varphi)^2}\right) dxdydzdr \quad . (24)$$

Here, we assume that the range variable $\tau$ is a discrete variable.

In this case, the preprocessing process also includes four operations but with slight difference. The first step of the preprocessing is a pulse compression. This pulse compression operation can be implemented by a convolution of the demodulated echo signal with a matched filter. For computational efficiency, this linear convolution operation is often implemented in the frequency domain by using FFT. After pulse compression, the signal in (24) becomes

$$S(t,\tau) = \iiiint_{r\,x,y,z} \text{sinc}\left(B_r\left(\tau - \frac{2r}{c}\right)\right) \cdot \exp\left\{-j\frac{4\pi}{c}f_c r\right\} \cdot g(x,y,z)$$
$$\cdot \delta\left(r - \sqrt{(x-R\cos\varphi\sin\theta)^2 + (y-R\cos\varphi\cos\theta)^2 + (z-R\sin\varphi)^2}\right) dxdydzdr \quad . (25)$$

The second step is a transformation of the range time variable $\tau$, which can be expressed as a change-of-variable defined as (13). For discrete signal, this transformation can be implemented by a resampling operation. As illustrated in Fig.4, the horizontal axis is the original variable $\tau$ and the vertical axis is the transformed variable $\tau'$, the mapping relationship between the two variables is shown by the curved line determined by (13). For an AD converter with constant sampling rate, the original sample positions on $\tau$ are uniformly distributed in the horizontal axis. After mapping into the new variable, the sample positions in the vertical axis become non-equally spaced. However, it is preferable to operate on a uniformly sampled data in the transformed variable in terms of computational efficiency. To this end, a resampling operation is performed on the data. As illustrated in Fig.4, the desired sampling positions on the new coordinate are predetermined with equally spaced squares in the vertical axis, then the resampling positions in the original variable, which are denoted by squares in the horizontal axis, can be determined by the mapping $\tau = \zeta(\tau')$.

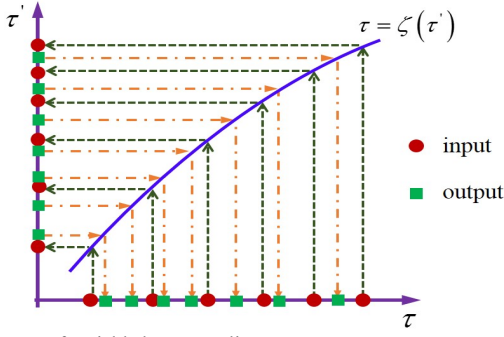

**Fig. 4.** Change of variable by resampling.

After range resampling, the signal becomes

$$S(t,\tau') = \int_u \iiint_{x,y,z} \mathrm{sinc}\!\left(\bar{B}_r\!\left(\tau' - \frac{2u}{c}\right)\right) \cdot \exp\!\left\{-j\frac{4\pi}{c}f_c r\right\} \cdot g(x,y,z)$$
$$\cdot \delta(u - x\cos\varphi\sin\theta - y\cos\varphi\cos\theta - z\sin\varphi)\,dxdydzdu \quad (26)$$

Because this change-of-variable operation is performed on the demodulated signal, the signal expression after transformation is slightly different from (17) which is the processing result operated on the modulated radio frequency signal. By comparing (26) with (17), to obtain the same result in (17), a phase adjustment is required on (26). The phase correction function is

$$H_1(u) = \exp\!\left\{j\frac{4\pi}{c}(f_c r - \bar{f}_c u)\right\}. \quad (27)$$

After this phase correction processing, the signal becomes

$$S(t,\tau') = \int_u \iiint_{x,y,z} g(x,y,z) \cdot \mathrm{sinc}\!\left(\bar{B}_r\!\left(\tau' - \frac{2u}{c}\right)\right) \cdot \exp\!\left\{-j\frac{4\pi}{c}\bar{f}_c u\right\}$$
$$\cdot \delta(u - x\cos\varphi\sin\theta - y\cos\varphi\cos\theta - z\sin\varphi)\,dxdydzdu \quad (28)$$

Equation (28) can also be expressed as

$$S(t,\tau') = \iiint_{x,y,z} g(x,y,z)\cdot\mathrm{sinc}\!\left(\bar{B}_r\!\left(\tau' - \frac{2(x\cos\varphi\sin\theta + y\cos\varphi\cos\theta + z\sin\varphi)}{c}\right)\right)$$
$$\cdot \exp\!\left\{-j\frac{4\pi}{c}\bar{f}_c(x\cos\varphi\sin\theta + y\cos\varphi\cos\theta + z\sin\varphi)\right\}dxdydz \quad (29)$$

The final preprocessing step is a range Fourier transform, which results in

$$S(t,f_\tau) = \iiint_{x,y,z} g(x,y,z)\cdot \exp\!\left\{-j\frac{4\pi}{c}(\bar{f}_c + f_\tau)(x\cos\varphi\sin\theta + y\cos\varphi\cos\theta + z\sin\varphi)\right\}dxdydz$$
$$f_\tau \in [-\bar{B}_r/2, \bar{B}_r/2] \quad (30)$$

To facilitate the image formation processing, the orbit plane is chosen as the XY plane such that the angle $\varphi = 0$. In this case, (30) can be simplified as

$$S(t,f_\tau) = \iiint_{x,y,z} g(x,y,z)\cdot\exp\!\left\{-j\frac{4\pi}{c}(\bar{f}_c + f_\tau)(x\sin\theta + y\cos\theta)\right\}dxdydz. \quad (31)$$

In actual, we only need to take into account the reflectivity of the Earth surface, then the target function can be modeled as

$$g(x,y,z) = r(x,y)\cdot\delta\!\left(z - \sqrt{R_0^2 - x^2 - y^2}\right), \quad (32)$$

where $r(x,y)$ is the surface reflectivity function at the orbit plane coordinates $(x,y)$, i.e., it is the projection of the illuminated Earth surface function $g(x,y,z)$ onto the orbit plane. Therefore, to reconstruct the target function $g(x,y,z)$ can be equivalent to reconstruct its projection.

Inserting (32) into (31) yields

$$S(t,f_\tau) = \iint r(x,y)\cdot\exp\!\left\{-j\frac{4\pi}{c}(\bar{f}_c + f_\tau)(x\sin\theta + y\cos\theta)\right\}dxdy. \quad (33)$$

This is the 2-D Fourier spectrum of the projection function. Due to the discrete property of the two independent variables, the obtained Fourier samples are spaced on a polar grid with polar radius $k_r = \frac{4\pi}{c}(\bar{f}_c + f_\tau)$ and angle $\theta$.

### B. Polar-To-Rectangular Reformatting

With the 2-D Fourier domain data, in order to reconstruct the target, the most straightforward way is to perform a 2-D inverse Fourier transform on the Fourier data. When dealing with the sampled data, the 2-D fast Fourier transform (FFT) algorithm is often a popular choice for an efficient implementation of the Fourier transformation. However, the 2-D FFT implementation requires uniformly spaced samples on a rectangular grid. Thus, for the actual polar-gridded data, a polar-to-rectangular resampling can be used to interpolate the 2-D polar-gridded data to a rectangular-gridded samples for ultimate use in the following 2-D FFT process. This polar reformatting is essentially a 2-D interpolation process, but a direct 2-D interpolation implementation on the 2-D data is still computationally intensive. To improve the computational efficiency, two separable 1-D interpolations, i.e., range resampling and azimuth resampling, are performed in the range dimension and azimuth dimension, respectively.

From the perspective of analytical analysis, the polar reformatting can also be interpreted as a decoupling procedure. Before polar reformatting, the two variables $t$ and $f_\tau$ in both phase terms in (33) are coupled together. After polar reformatting, it is desired that the coefficient of the range coordinate $y$ in (33) becomes a univariate linear function of the range frequency variable and the coefficient of the azimuth coordinate $x$ becomes a univariate linear function of the azimuth variable. Specifically, the range resampling is dedicated to eliminating the coupling between range and azimuth variables in the coefficient of $y$, making the coefficient only a univariate linear function of range frequency variable. This procedure can be implemented by performing an angle-dependent change-of-variable on range frequency in (30), i.e., setting $f_\tau = \vartheta_r(\tilde{f}_\tau; t)$, where $\tilde{f}_\tau$ is the new range frequency variable. This change-of-variable should make sure that the following decoupling transformation is achieved:

$$\frac{4\pi}{c}(\bar{f}_c + f_\tau)(x\sin\theta + y\cos\theta) \xrightarrow{f_\tau = \vartheta_r(\tilde{f}_\tau; t)} \frac{4\pi}{c}(f_c + \tilde{f}_\tau)(x\tan\theta + y), \quad (34)$$

From (34), we can easily get

$$\vartheta_r(\tilde{f}_\tau; t) = \delta_r \tilde{f}_\tau + \bar{f}_c(\delta_r - 1), \quad (35)$$

where $\delta_r = 1/\cos\theta$. Equation (35) shows that the range resampling is essentially a range frequency scaling transformation with an offset $\bar{f}_c(\delta_r - 1)$.

Undergoing the above change-of-variable, the signal in (33)

becomes

$$S_R(t, \tilde{f}_\tau) = S[t, \vartheta_r(\tilde{f}_\tau; t)] = \iint_{x,y} r(x,y) \cdot \exp\left\{-j\frac{4\pi}{c}(\bar{f}_c + \tilde{f}_\tau)(x\tan\theta + y)\right\} dxdy. \quad (36)$$

In (36), there has been no coupling between the new range frequency $\tilde{f}_\tau$ and the azimuth angle $t$ in the coefficient of $y$ term.

Similarly to the range resampling, the azimuth resampling is dedicated to eliminating the coupling between the azimuth variable and the range variable in the coefficient of $x$ in (36), making the coefficient only a univariate linear function of the azimuth variable. This procedure can be implemented by performing a range-frequency-dependent change-of-variable on azimuth time variable, which can be denoted as $t = \vartheta_a(\tilde{t}; \tilde{f}_\tau)$. After this change-of-variable, the signal becomes

$$S_A(\tilde{t}, \tilde{f}_\tau) = S_R[\vartheta_a(\tilde{t}; \tilde{f}_\tau), \tilde{f}_\tau]$$
$$= \iint_{x,y} r(x,y) \cdot \exp\left\{-j\left[\frac{4\pi \bar{f}_c \Omega}{c}\tilde{t}x + \frac{4\pi}{c}(\bar{f}_c + \tilde{f}_\tau)y\right]\right\} dxdy. \quad (37)$$

Now the coefficient of $x$ term is a univariate linear function of the new azimuth variable $\tilde{t}$, and the coefficient of $y$ term is a univariate linear function of the new range variable $\tilde{f}_\tau$. By defining $k_x = \frac{4\pi \bar{f}_c \Omega}{c}\tilde{t}$, $k_y = \frac{4\pi}{c}(\bar{f}_c + \tilde{f}_\tau)$, (37) can be rewritten as

$$S_A(k_x, k_y) = \iint_{x,y} r(x,y) \cdot \exp\{-j(k_x x + k_y y)\} dxdy. \quad (38)$$

Because the samples on $\tilde{t}$ and $\tilde{f}_\tau$ domain are uniformly spaced, after mapping to the $k_x$ and $k_y$ domain by the linear transformation, the samples are also uniformly spaced. This rectangular-gridded 2-D sample data provide an appropriate form for the efficient implementation of Fourier inversion by FFT.

### C. 2-D IFFT

After the above polar reformatting, the residual signal is a 2-D sinusoid signal. Therefore, a 2-D inverse Fourier transform can produce a focused image of the target. Because the samples are now uniformly spaced on a rectangular grid, a 2-D IFFT can be applied as a fast implementation of the inverse Fourier transform.

$$\hat{r}(x,y) = \mathbf{F}^{-1}[S_A(k_x, k_y)]$$
$$= \iint_D \left\{\iint r(x',y') \cdot \exp\{-j(k_x x' + k_y y')\} dx'dy'\right\} \cdot \exp\{j(k_x x + k_y y)\} dk_x dk_y, \quad (39)$$

where $\mathbf{F}^{-1}[\cdot]$ represents the 2-D inverse Fourier transform, $D$ is the signal support area in the wavenumber domain.

Exchanging the integration order, (39) can also be written as

$$\hat{r}(x,y) = \iint \left\{r(x',y') \cdot \iint_D \exp\{j[k_x(x-x') + k_y(x-x')]\} dk_x dk_y\right\} dx'dy', \quad (40)$$
$$= r(x,y) \otimes \text{psf}(x,y)$$

where the symbol $\otimes$ represents the convolution operation, $\text{psf}(x,y)$ is the 2-D point spread function (PSF) defined as

$$\text{psf}(x,y) = \iint_D \exp\{j(k_x x + k_y y)\} dk_x dk_y. \quad (41)$$

If the support area is the whole 2-D wavenumber domain plane, the PSF will be the 2-D Dirac delta function, then the projected target function $r(x,y)$ can be accurately reconstructed.

However, in actual, due to the limited transmission bandwidth and viewing angle, the obtained spectrum support area is limited in a polar annular region. This area can often be approximated as a rectangle. If we assume that the widths of the support area in $k_x$ and $k_y$ direction are $\Delta k_x$ and $\Delta k_y$, respectively, then the PSF can be approximated as a 2-D sinc function

$$\text{psf}(x,y) = \text{sinc}\left(\frac{\Delta k_x}{2\pi}x\right) \cdot \text{sinc}\left(\frac{\Delta k_y}{2\pi}y\right), \quad (42)$$

where $\text{sinc}(\cdot)$ is the sinc function defined as $\text{sinc}(u) = \frac{\sin \pi u}{\pi u}$.

## IV. SPHERICAL GEOMETRY ALGORITHM FOR HIGH-RESOLUTION SAR

As the SAR resolution increases, the required synthetic aperture time becomes very large. The increased synthetic aperture time will complicate the image formation processing from at least two aspects. First, the flight trajectory of the radar platform can not be approximated as a straight line, or even a circular arc, but must be modeled as an elliptical arc. Secondly, the effect of the Earth's rotation can't be ignored anymore. Although the satellite orbit is a plannar curve line in the ECI coordinate system, the trajectory in the ECEF coordinate system, which is the most widely used frame in the image formation process, will becomes a nonplanar curve line (curvilinear) when taking into account the Earth's rotation. The first effect, i.e., the plannar nonlinear trajectory, can be automatically corrected during the proposed image formation processing by adjusting the phase compensation term and the input parameters for the range and azimuth resampling. However, the nonplannar motion will introduce additional space-variant phase error which can not be compensated automatically by the image formation algorithm proposed in the last section. In this case, phase history data are collected on a non-planar ribbon in 3-D Fourier space. Treating this non-plannar Fourier ribbon as if it were a plannar surface will induce phase error because of the distorted Fourier space. The subsequent image formation processing will produce a defocused imagery. Therefore, an out-of-plane correction is required to produce a well-focused image.

In this section, we will provide a detailed analysis on the nonplannar motion effect and propose an efficient approach to correct for the undesired phase error introduced by the nonplannar motion.

### A. Earth's Rotation and Nonplanar Trajectory Effect

In satellite orbit dynamics, ECI coordinate frame is often the most commonly used coordinate system, because the satellite orbit in this frame is a planar circle or ellipse. However, in the SAR image formation processing, what we are concerned about is the relative geometric relationship between the satellite and the illuminated Earth surface. Therefore, the ECEF coordinate frame will be a more appropriate choice during the image formation processing. In the ECEF frame, because of the rotation of the Earth, the satellite orbit becomes a more complicated nonplanar curve. The out-of-plane motion (variation of the grazing angle $\varphi$) brings two adverse effects on the phase history data, which can be clearly seen in (19). For convenience, we rewrite the formula as follows

$$S(t,f_\tau) = \iiint_{x,y,z} g(x,y,z) \cdot \exp\left\{-j\frac{4\pi}{c}(\bar{f}_c + f_\tau)\right. \\ \left. \cdot (x\cos\varphi\sin\theta + y\cos\varphi\cos\theta + z\sin\varphi)\right\} dxdydz \quad (43)$$

Firstly, the third exponential term in (43), which is an undesired phase error term during the 2-D image formation processing, can not be ignored anymore. From (43), it is clear that there are two cases where the third exponential term is negligible, one is $z=0$ and the other one is $\varphi=0$. In conventional SAR processing, the ground plane can be chosen as the focus plane, then the z coordinates of all the scatterers in this coordinate system equal zero. In another case, if the radar platform trajectory during the integration time is a planar curve, an appropriate coordinate system can be chosen to ensure that $\varphi=0$. However, for the proposed algorithm for high resolution SAR, neither of the above conditions can be met. On the one hand, the Earth center is used as the origin of the coordinate frame in the proposed algorithm, therefore, it is impossible to select a focus plane that coincides with the illuminated ground plane to ensure that $z=0$ for all the scatterers. On the other hand, the out of planar motion relative to the illuminated ground surface induces a continuous change of the grazing angle $\varphi$. This will lead to the fact that there is no case where $\varphi=0$ exists. Because of these reasons, to produce a well-focused imagery, the third exponential term in (43) must be eliminated during the image formation processing.

Even if the above phase error term has been corrected, the nonplanar motion still has another adverse effect. Because the grazing angle $\varphi$ is not equal to zero and is constantly changing with azimuth time, the phase history data are collected on a non-planar ribbon in the 3-D Fourier space. Treating this non-plannar Fourier ribbon as if it were a plannar surface will induce phase error because of the distorted Fourier space.

B. *Phase Error Compensation*

From (43), it is clear that the 2-D phase error in the phase history domain can be expressed as

$$\Phi_e(t,f_\tau) = -\frac{4\pi}{c}(\bar{f}_c + f_\tau) \cdot z \cdot \sin[\varphi(t)]. \quad (44)$$

Because all the scatterers are located on the ground surface, then the z coordinates satisfy $z = \sqrt{R_0^2 - x^2 - y^2}$. Therefore, the phase error can also be expressed as

$$\Phi_e(t,f_\tau) = -\frac{4\pi}{c}(\bar{f}_c + f_\tau) \cdot \sqrt{R_0^2 - x^2 - y^2} \sin[\varphi(t)]. \quad (45)$$

From the formula, it is clear that the 2-D phase error is space-variant, i.e., the phase errors are different for scatterers located at different positions. Furthermore, the space-variant effect appears in both range and azimuth dimensions. To quantitatively evaluate the space-variance effect, the first derivatives of the phase error with respect to the azimuth and range variables at the scene center are computed

$$\frac{\partial}{\partial x}\Phi_e(t,f_\tau)\bigg|_{x=x_c,y=y_c} = \frac{4\pi}{c}(\bar{f}_c + f_\tau)\cdot\sin[\varphi(t)]\frac{x_c}{\sqrt{R_0^2 - x_c^2 - y_c^2}} \\ \frac{\partial}{\partial y}\Phi_e(t,f_\tau)\bigg|_{x=x_c,y=y_c} = \frac{4\pi}{c}(\bar{f}_c + f_\tau)\cdot\sin[\varphi(t)]\frac{y_c}{\sqrt{R_0^2 - x_c^2 - y_c^2}}, \quad (46)$$

where $(x_c, y_c)$ is the coordinate of the scene center.

For radar operated in broadside mode, the azimuth coordinate $x_c$ is approximately equal to zero, but range coordinate $y_c$ has a large offset. Therefore, the first derivative of the phase error along the azimuth dimension is very small and the space-variant effect along this dimension is negligible, but the first derivative along the range dimension is relatively large thus the range space-variant effect of the phase error should be taken into account.

To eliminate the space-variant phase error, the compensation procedure is divided into two steps, one is space-invariant compensation with respect to the scene center, the other one is range-dependent space-variant compensation. The space-invariant phase error correction can be performed in the phase history domain by a conjugate multiplication of the 2-D phase error in the scene center

$$H_2(t,f_\tau) = \exp\left\{\frac{4\pi}{c}(\bar{f}_c + f_\tau) \cdot \sqrt{R_0^2 - x_c^2 - y_c^2}\sin[\varphi(t)]\right\}. \quad (47)$$

This 2-D phase error compensation includes a range migration error correction and an azimuth phase error compensation. The range-dependent space-variant phase compensation should be performed after range compression because the range information is required during the correction procedure. To this end, this correction is performed after polar reformatting and range FFT. After the space-invariant compensation, the residual 2-D phase error for the scatterer located at the range coordinate y is

$$\Phi_{PH}(t,f_\tau) = -\frac{4\pi}{c}(\bar{f}_c + f_\tau)\cdot\left(\sqrt{R_0^2 - x^2 - y^2} - \sqrt{R_0^2 - x_c^2 - y_c^2}\right)\cdot\sin[\varphi(t)]. \quad (48)$$

Because the azimuth space-variant effect is negligible, this residual phase error can be approximated as

$$\Phi_{PH}(t,f_\tau) \approx -\frac{4\pi}{c}(\bar{f}_c + f_\tau)\cdot\left(\sqrt{R_0^2 - x_c^2 - y^2} - \sqrt{R_0^2 - x_c^2 - y_c^2}\right)\cdot\sin[\varphi(t)]. \quad (49)$$

It should be noted that this is the residual phase error in the phase history domain, but now what we are concerned about is the residual phase error after polar reformatting because the space-variant compensation should be performed after polar transformation.

From the previous section, we know that the polar reformatting includes a range resampling and an azimuth resampling. The range resampling is a range scaling transformation denoted as $f_\tau = \vartheta_r(\tilde{f}_\tau;t)$, then the residual phase error after range resampling can be expressed as

$$\Phi_{RR}(t,\tilde{f}_\tau) = \Phi_{PH}(t,\vartheta_r(\tilde{f}_\tau;t)) = -\frac{4\pi(\bar{f}_c + \tilde{f}_\tau)}{c}\cdot\left(\sqrt{R_0^2 - x_c^2 - y^2} - \sqrt{R_0^2 - x_c^2 - y_c^2}\right)\cdot\frac{\tan[\varphi(t)]}{\cos[\theta(t)]}. \quad (50)$$

The azimuth resampling is range-dependent azimuth time transformation denoted as $t = \vartheta_a(\tilde{t}; \tilde{f}_\tau)$, then the residual phase error after azimuth resampling is expressed as

$$\Phi_{AR}(\tilde{t},\tilde{f}_\tau) = \Phi_{RR}(\vartheta_a(\tilde{t};\tilde{f}_\tau),\tilde{f}_\tau)$$
$$= -\frac{4\pi(\bar{f}_c + \tilde{f}_\tau)}{c} \cdot \left(\sqrt{R_0^2 - x_c^2 - y^2} - \sqrt{R_0^2 - x_c^2 - y_c^2}\right) \cdot \frac{\tan[\varphi(\vartheta_a(\tilde{t};\tilde{f}_\tau))]}{\cos[\theta(\vartheta_a(\tilde{t};\tilde{f}_\tau))]}. \quad (51)$$

In essence, it still includes an azimuth phase error term and a residual range migration term. But the residual range migration term after space-invariant compensation is often very small and can be negligible, then only the azimuth phase error term

$$\Phi_{AR}(\tilde{t}) \approx -\frac{4\pi \bar{f}_c}{c} \cdot \left(\sqrt{R_0^2 - x_c^2 - y^2} - \sqrt{R_0^2 - x_c^2 - y_c^2}\right) \cdot \frac{\tan[\varphi(\vartheta_a(\tilde{t};0))]}{\cos[\theta(\vartheta_a(\tilde{t};0))]}, \quad (52)$$

needs to be compensated.

Therefore, the space-variant phase compensation involves multiplying the range-compressed data (in the $\tilde{t} - y$ domain) by the complex conjugate of the residual azimuth phase error, i.e.,

$$H_3(\tilde{t},y) = \frac{4\pi \bar{f}_c}{c} \cdot \left(\sqrt{R_0^2 - x_c^2 - y^2} - \sqrt{R_0^2 - x_c^2 - y_c^2}\right) \cdot \frac{\tan[\varphi(\vartheta_a(\tilde{t};0))]}{\cos[\theta(\vartheta_a(\tilde{t};0))]}. \quad (53)$$

### C. Projection of the 3-D Ribbon Data onto Plane

From the previous section we have known that the phase history domain data after some preprocessing are collection of 3-D complex sinusoids in the Fourier space of the illuminated 3-D scatterers. For any planar slice of the Fourier space, a high-efficient Fourier inversion processing can produce a focused imagery for all the illuminated scatterers. This 2-D image is the projection of the illuminated 3-D object normal to the Fourier slice.

However, for a very high resolution synthetic aperture radar, due to the non-planar data collection in physical space, the phase history data are collected on a non-planar ribbon in the 3-D Fourier space. Treating this curved Fourier surface as if it were a planar one will induce additional phase errors and result in poor focused imagery. Therefore, to obtain a well-focused image, an out-of-plane compensation is a necessary procedure.

Fortunately, after the above phase error compensation, the z-coordinate dependent phase term is completely eliminated. In this case, the Fourier data does not vary along the $k_z$ direction, then the 3-D Fourier data can be projected onto the $k_z = 0$ plane without any approximation. This projection process can be embedded in the range interpolation process of polar reformatting, so no additional computation is needed. In practice, the only adjustment is the input parameter for the range resampling. Next, we will give a theoretical derivation to show this modification.

The phase history domain signal after phase error compensation can be expressed as

$$S(t,f_\tau) = \iint_{x,y} r(x,y) \cdot \exp\left\{-j\frac{4\pi}{c}(\bar{f}_c + f_\tau)(x\cos\varphi\sin\theta + y\cos\varphi\cos\theta)\right\} dxdy. \quad (54)$$

As show in the previous section, the range resampling in polar reformatting is to eliminate the coupling between the range and azimuth variables in the coefficient of $y$, making the coefficient only an univariate linear function of the range frequency variable. To this end, the range resampling can be expressed as the following scaling transform

$$f_\tau = \vartheta_r(\tilde{f}_\tau; t) = \delta_r \tilde{f}_\tau + \bar{f}_c(\delta_r - 1), \quad (55)$$

where $\delta_r = 1/(\cos\theta\cos\varphi)$. This scaling transform is very similar with the one shown in (35), the only difference is the value of scaling factor $\delta_r$.

After the range resampling, the signal becomes

$$S_{RR}(t,\tilde{f}_\tau) = S(t,\vartheta_r(\tilde{f}_\tau;t))$$
$$= \iint_{x,y} r(x,y) \cdot \exp\left\{-j\frac{4\pi}{c}(\bar{f}_c + \tilde{f}_\tau)(x\tan\theta + y)\right\} dxdy. \quad (56)$$

This expression is exactly the same as (36). Therefore, the followed image formation steps, including an azimuth resampling and an 2-D Fourier transform, are exactly the same as the approach proposed in the previous section.

### D. Processing Flow

Incorporating the above nonplanar motion compensation, a modified spherical geometry algorithm for high resolution spaceborne SAR is proposed, whose processing flow is shown in Fig.5. Compared with the low resolution spherical geometry algorithm, there are three differences. First, a first-order motion compensation is added, which includes a space-invariant APE compensation and residual RCM correction performed in the phase history domain (after range preprocessing). Secondly, after polar reformatting and range compression, a second-order motion compensation, i.e., a range-dependent residual APE correction, is embedded to eliminate the residual space-variant phase error. Thirdly, a data projection processing is applied to facilitate the fast implementation of Fourier inversion. This process is integrated into the range resample procedure in polar reformatting, therefore, no extra computation is required.

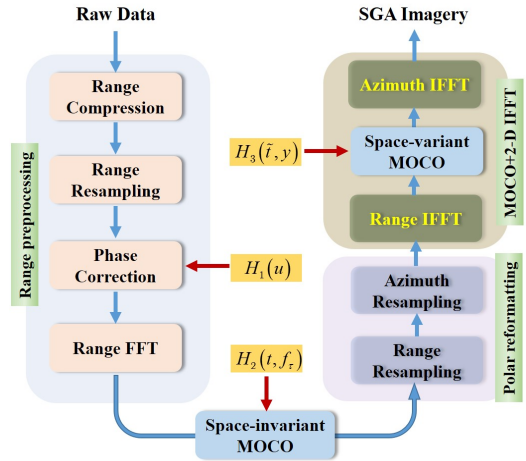

**Fig. 5.** Processing flow of the proposed SGA.

## V. EXPERIMENTAL RESULTS

In this section, both numerical simulations and real data processing are performed to validate the effectiveness of the proposed approach. First, to assess the focus quality in the cases of ultra-high resolution and very large scene, two data sets with different system parameters are simulated, one is for very large scene but with relatively low resolution and the other is for ultra-high resolution but with relatively small scene. The reason why not choose a data set with both very large scene and very high resolution is just due to memory limit. The radar operates in spotlight mode, some main radar parameters are shown in Table I.

**Table I  Simulation Parameters**

| Parameter | Wide Swath | High Resolution |
| --- | --- | --- |
| Carrier Frequency | 5GHz | 10GHz |
| Range Bandwidth | 30MHz | 1.5GHz |
| Pulse Duration | 5us | 2us |
| Coherent Time | 3s | 15s |
| Orbit Altitude | 2000km | 600km |
| Reference Range | 3500km | 760km |
| Nominal Resolution | 5m*5m | 0.1m*0.1m |
| Scene Size | 300km*300km | 10km*10km |
| Raw Data Size | 64K*128K | 128K*128K |

In order to facilitate the evaluation of the focusing performance of the algorithm, six typical point targets located at different positions in the scene are assumed. All the scatterers are located on the spherical surface of the Earth, and their geometric relationship with respect to the radar is shown in Fig.6. The 3-D coordinate distributions of these point targets are shown in Fig.6(a) and Fig.6(b), respectively, for two data set. In the wide scene mode, the point target position corresponds to a scene size of 300km, while in the high-resolution mode it corresponds to 10km.

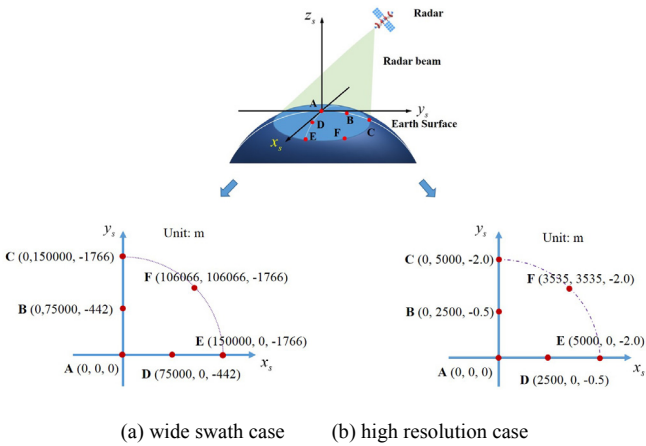

(a) wide swath case      (b) high resolution case

**Fig. 6.** Data collection geometry and point target distribution on the spherical ground surface.

*A. Wide-Swath Data*

For the wide swath data, the radar operates on a MEO satellite with an orbit height of about 2000km. The instantaneous position and velocity data of the simulated satellite are produced by STK software. The simulated synthetic aperture time is about 3s, corresponding to a theoretical azimuth resolution of 5m. During this coherent time, the instantaneous 3-D coordinates of the radar in the rotated ECEF frame is shown in Fig.7. Because the Z coordinate is small enough during the coherent pulse interval, the flight trajectory of the radar platform can be approximated as a plane curve in the X-Y plane, therefore the out-of-plane motion effect can be ignored.

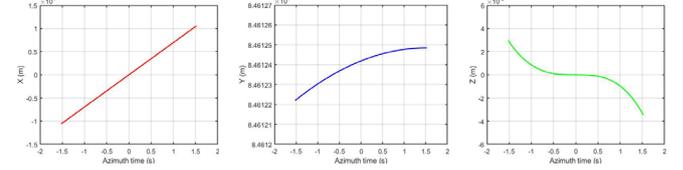

**Fig. 7.** The instantaneous 3-D coordinates of the radar in the rotated ECEF frame.

To verify the algorithm's inherent compensation capability for the spherical ground surface effect, the simulated point targets are assumed to distribute on the spherical ground surface. To increase the influence of the spherical surface, a very large illuminated scene of size 300km*300km is simulated in this case.

The simulated data is processed by the proposed low-resolution SGA algorithm, and the 2-D contour plots of the six focused targets are shown in Fig.8. It can be seen that all the point targets are focused well. For comparison, the data is also processed by PFA and FBP. The azimuth profiles of the six point targets produced by the three different algorithms are shown in Fig. 9. Although PFA algorithm has very excellent nonlinear trajectory compensation capability, it also has two obvious shortcomings. First, PFA algorithm has a plane wavefront assumption, which limits the size of the imaged scene, e.g, the focus radius of PFA in the simulated scenario is only about 76km, which is less than the expected 150km. Furthermore, in analogy to other frequency domain algorithm, the PFA also has a flat ground surface assumption. For the very large illuminated scene, the out-of-plane effect for the targets far away from the scene center will become nonnegligible. This can be clearly seen in the PFA imagery shown in Fig.9, where the point target located in the scene center is well focused, but the targets far away from the scene center begin to suffer from image degradation. As a contrast, the proposed SGA has no approximation both on spherical wavefront and spherical ground surface, therefore, it can provide excellent focus quality for all the targets. This can be clearly seen from the azimuth profile of the point target response in Fig.9, which show that SGA has almost the same focus capability as FBP algorithm.

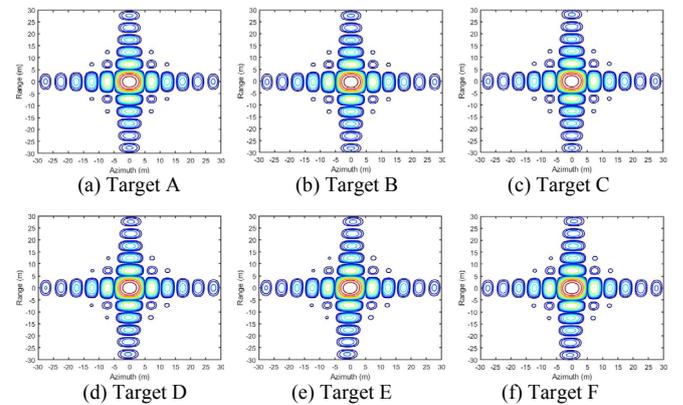

**Fig. 8.** Processing results by SGA

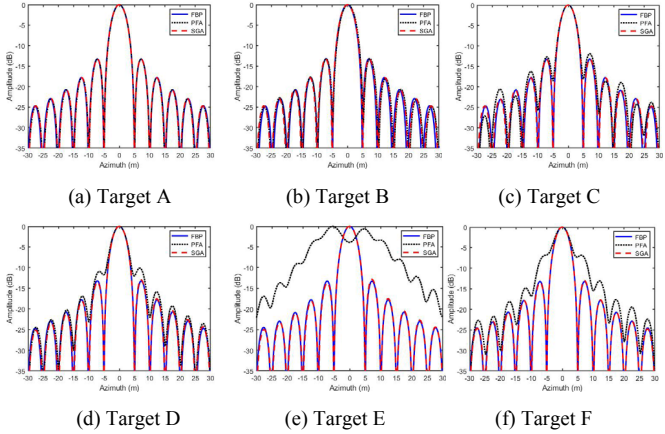

(a) Target A  (b) Target B  (c) Target C

(d) Target D  (e) Target E  (f) Target F

**Fig. 9.** Azimuth profile of point target response for different algorithms.

*B. Ultra High Resolution Data*

While for the high resolution data, the radar operates on a LEO satellite with an orbit height of 500km. The simulated synthetic aperture time is 15 seconds, corresponding to a theoretical azimuth resolution of 0.1m. During this very long synthetic aperture time, the instantaneous position of the radar platform in the rotated ECEF frame is shown in Fig.10. From the figure, we can clearly see the out-of-plane motion effect due to the Earth's rotation.

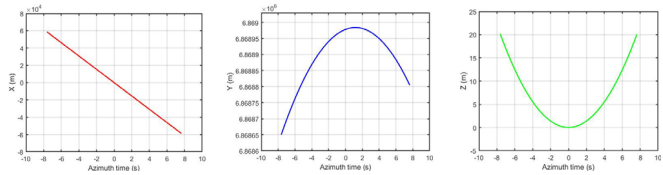

**Fig. 10.** The instantaneous 3-D coordinates of the radar in the rotated ECEF frame.

First, the PFA algorithm is applied on the data set and the produced imagery is shown in Fig.11. In this scenario, the spherical ground surface effect becomes negligible due to limited scene size. However, because of the plannar wavefront approximation, the focus radius of PFA in this simulation condition is only 710m, which is far less than the simulated scene size. Therefore, in the PFA imagery shown in Fig.11, the targets with a distance of more than the focus radius from the scene center suffer from serious defocus.

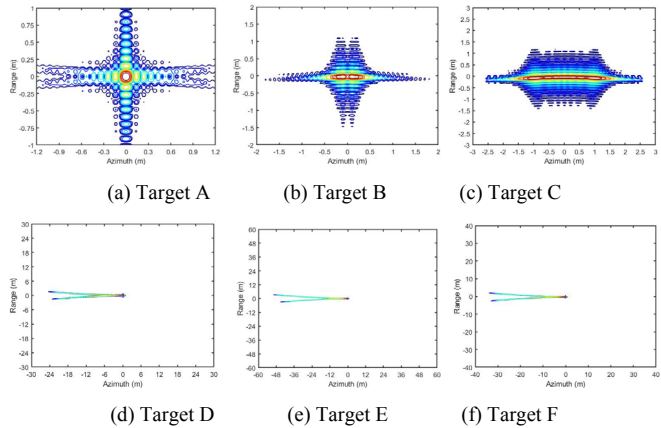

(a) Target A  (b) Target B  (c) Target C

(d) Target D  (e) Target E  (f) Target F

**Fig. 11.** Processing results by PFA.

Then, the simulated high resolution data is processed by the proposed high resolution SGA algorithm and the results is shown in Fig.12. The interpolated 2-D contour plot of the point spread function of the six targets clearly show that all the targets in the imagery produced by the proposed algorithm are well focused. Furthermore, all the measured focus quality indexes, including impulse response width, peak sidelobe ratio and integrated sidelobe ratio, are all very close to the theoretical values.

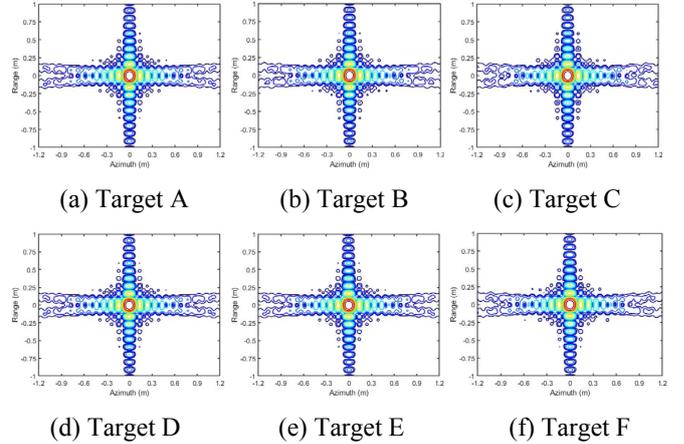

(a) Target A  (b) Target B  (c) Target C

(d) Target D  (e) Target E  (f) Target F

**Fig. 12.** Processing results by SGA.

*C. Real Data Results and Analysis*

We have also successfully processed several on-board spaceborne SAR real data using the proposed algorithm, and the results for one of them are shown here. The used real data are collected by a Chinese commercial spaceborne SAR Chaohu-1. The raw data are collected by a C-band radar operating in spotlight mode with a standoff range of 630km. The bandwidth of the transmitted radar signal is 300MHz, corresponding to a theoretical slant range resolution of 0.5m. The processed synthetic aperture length is 34km. Therefore, the theoretical azimuth resolution is also 0.5m. First, the data is processed by PFA and the produced imagery is shown in Fig.13. The horizontal axis of the PFA image is the range direction and the range increases from left to right. The scene dimension of the whole image is about 8km*14km (azimuth*range). At first glance, the image appears to be well-focused, but a closer look at the local area, as shown in Fig.15(a), shows that only the central area of the scene is in full focus, while the targets at the edges of the scene still suffer from defocus.

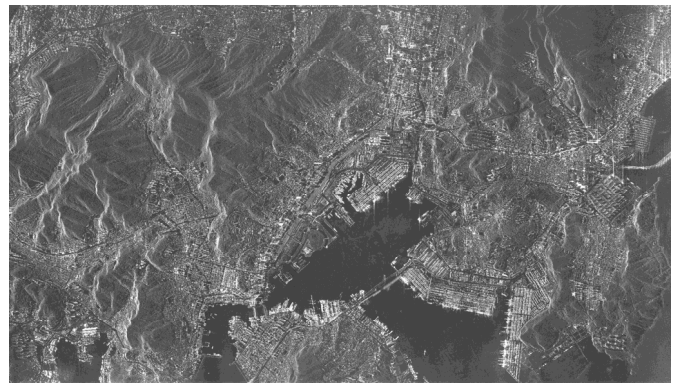

**Fig. 13.** Spotlight SAR imagery produced by PFA.

Fig. 14 shows the processing result of the proposed SGA

method, and similarly, in order to see more clearly the focusing quality of the local details, the magnified image of the local region is given in Fig. 15(b). It can be seen that after processing by this method, different areas in the scene are all well focused.

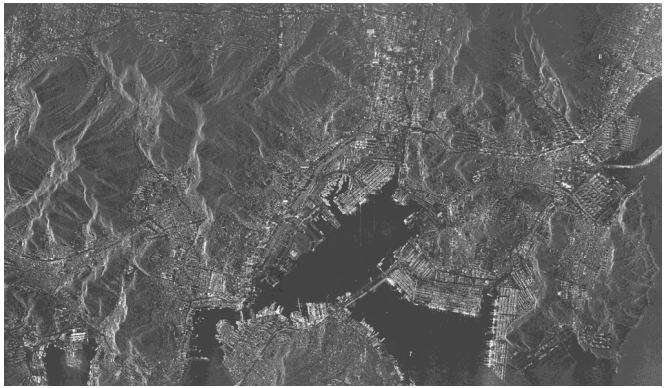

**Fig. 14.** Spotlight SAR imagery produced by SGA.

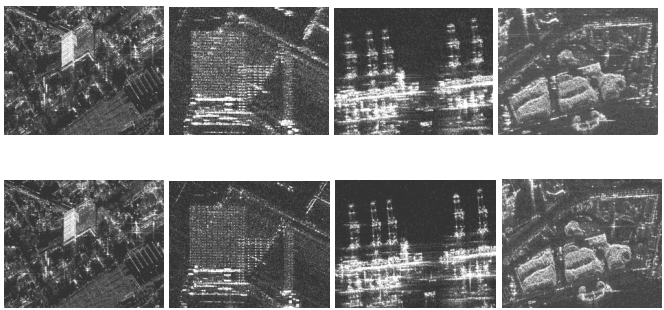

**Fig. 15.** Enlarged local areas in Fig. 13 and Fig.14.

VI. CONCLUSION

This paper has presented a new image formation algorithm, named SGA, which is specifically designed for spaceborne synthetic aperture radar data processing. The new approach utilizes the unique spherical geometry property of the ground surface in spaceborne SAR imaging to establish the exact Fourier transform relationship between the radar echo signal and the target function. It then implements FFT-based image reconstruction from the collected non-planar Fourier space data by some non-planar effect compensation operations such as data projection and space-variant phase correction. The SGA algorithm can provide efficient and accurate image formation processing even under very complicated imaging conditions, such as nonlinear radar flight trajectory, and spherical ground surface. Therefore, it has great potential for application in high-resolution and/or wide-swath spaceborne SAR imaging. Both simulated and measured data processing results fully validate the effectiveness of the proposed algorithm.

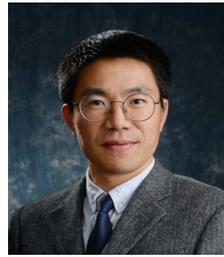

**Xinhua Mao** (Member, IEEE) was born in Hunan, China. He received the B.S. and Ph.D. degrees from the Nanjing University of Aeronautics and Astronautics (NUAA), Nanjing, China, in 2003 and 2009, respectively, all in electronic engineering.

In 2009, he joined the Department of Electronic Engineering, NUAA, where he is currently a Full Professor with the Key Laboratory of Radar Imaging and Microwave Photonics. He was a Visiting Scholar with Villanova University, Villanova, PA, USA, in 2013, and the University of Leicester, Leicester, U.K., from 2018 to 2019. His research interests include radar imaging, and ground moving target indication (GMTI), inverse problems. He has developed algorithms for several operational airborne SAR systems.

Dr. Mao was a recipient of the one National Science and Technology Progress Award in 2019 and three National Defense Technology Awards in 2007, 2015, and 2018, in China. He received the Best Paper Award in the 5th Asia-Pacific Synthetic Aperture Radar Conference in 2015.